# Interface mediated mechanisms of plastic strain recovery in a AgCu alloy

B.P. Eftink[1], A. Li[2], I. Szlufarska[2], and I.M. Robertson[2]

[1]Department of Materials Science and Engineering, University of Illinois, Urbana IL. 61801

[2]Department of Materials Science and Engineering, University of Wisconsin-Madison, Madison WI. 53706

**Abstract:**

Through the combination of transmission electron microscopy analysis of the deformed microstructure and molecular dynamics computer simulations of the deformation processes, the mechanisms of plastic strain recovery in bulk AgCu eutectic with either incoherent twin or cube-on-cube interfaces between the Ag and Cu layers and a bilayer thickness of 500 nm have been revealed.  The character of the incoherent twin interfaces changed uniquely after dynamic compressive loading for samples that exhibited plastic strain recovery and was found to drive the recovery which is due to dislocation retraction and rearrangement of the interfaces.  The magnitude of the recovery decreased with increasing strain as dislocation tangles and dislocation cell structures formed.  No change in the orientation relationship was found at cube-on-cube interfaces and these exhibited a lesser amount of plastic strain recovery in the simulations and none experimentally in samples with larger layer thicknesses with predominantly cube-on-cube interfaces.  Molecular dynamics computer simulations verified the importance of the change in the incoherent twin interface structure as it was found to be the driving force for dislocation annihilation at the interfaces and the plastic strain recovery.

## 1. Introduction:

High interface density materials have shown favorable properties in addition to high mechanical strength such as corrosion resistance, wear resistance, irradiation tolerance, and high electrical conductivity [1-6]. A fundamental understanding of the relationships between synthesis of materials with a high density of interfaces, the structure of those interfaces, and the resultant properties provides an avenue to guide material design. The AgCu eutectic system is a model material system for studying the dependence of the deformation mode and the transfer of strain across FCC/FCC interfaces as the interface can have either a cube-on-cube or coherent twin orientation relationship between the Ag and Cu layers with $\{111\}_{Ag}||\{111\}_{Cu}$ habit planes or incoherent twin interfaces with $\{\bar{3}13\}_{Ag}||\{\bar{1}12\}_{Cu}$ habit planes [7-9]. In directionally solidified material, the percentage of cube-on-cube versus incoherent twin interfaces decreases with decreasing bilayer thickness, which is controlled by processing conditions. The interface type has important implications on the overall mechanical response of the material as the interaction of dislocations, both perfect and partial, is dependent on it [7]. For example, it has been found that the cube-on-cube interface permits transmission of deformation twins across it whereas the incoherent and coherent twin interfaces do not [7, 10, 11]. Furthermore, in the 500 nm bilayer thick AgCu eutectic, which has a preponderance of incoherent twin interfaces, following loading parallel to the interfaces, i.e., along the $[101]_{Ag}||[110]_{Cu}$ growth direction, at a strain rate of $10^3$ s$^{-1}$ but not at $10^{-3}$ s$^{-1}$, plastic strain recovery occurred [12]. The magnitude of this plastic strain recovery decreased with increasing strain; 3.2 % plastic strain recovery was observed at a true strain of 9.3 % and no recovery was reported in samples loaded to a true strain of 21.9 %.  Kingstedt *et al*. reported that the degree of recovery decreased with a change in the loading orientation with respect to the interface normal [12].  For example, loading 90° to the $[101]_{Ag}||[110]_{Cu}$ growth direction at a strain rate of $10^3$ $s^{-1}$ to a true strain of 15.4 % resulted in a plastic strain recovery of 0.6 %.  Figure 1 shows the stress-strain curves for the 500 nm bilayer thick eutectic loaded at different orientations, the true strains applied by the split-Hopkinson pressure bar compression (SHPB) are marked by the closed symbols while the true strains measured post compression are marked by the open symbols [12].  The dependence of the magnitude of the plastic recovery on the loading direction with respect to the interface, and the strain is summarized in Table 1.  Kingstedt *et al*. proposed that a number of mechanisms and microstructural features acted collaboratively in the plastic recovery, although no definitive mechanism was identified [12].  They did, however, eliminate temperature increase as a possible mechanism as it was determined to be on the order of 10-12 degrees.  Herein the microstructural state of the materials whose response is shown in Figure 1 is determined.   Kingstedt *et al*. also reported that no measurable plastic strain recovery occurred as the bilayer thickness increased and the majority of interfaces transitioned to cube-on-cube [12].

Plastic strain recovery has been found experimentally following room temperature deformation to a strain of 2 % in Al and Au thin films with 65 and 50 nanometer grain sizes, respectively [13].  However, for the recovery in Au to be observed, the specimens had to be heated to 210 °C.  The amount of recovery in Al and Au was 0.33 % and 0.35 % strain, respectively  [13].  Al films with a grain size of 180 nm did not exhibit recovery, even after annealing for 30 min at 220 °C.  Molecular dynamics (MD) computer simulations have been performed to explain the recovery observed in Al and Au and it was attributed to dislocation retraction into the grain boundary along with grain boundary sliding [14].  Strain recovery has also been observed in a penta-twinned Ag nanowire strained *in situ* in a transmission electron microscope (TEM) [15]. The recovery was attributed to the leading partial dislocations, which had been blocked by the twin boundaries, retracting to the nucleation site of the free surface on unloading; this mechanism was supported by the results of MD simulations [15]. A MD study of the deformation processes in a multilayered system comprised of both 2 and 8 nm-thick Cu and Ag layers with a cube-on-cube orientation relationship and $(111)_{Ag}||(111)_{Cu}$ interface habit planes showed that straining in tension along [111] to a strain of 4% was carried by deformation twinning [16]. Complete recovery of the plastic strain was reported on unloading. However, since volume relaxation along the two spatial directions perpendicular to the loading axis was not allowed, the simulations did not account for the Poisson effect and it is unclear if such a constraint on volume could influence the plastic strain recovery mechanism. Furthermore, a measurable level of plastic strain recovery was not observed experimentally in a AgCu eutectic alloy in which the majority of the interfaces were cube-on-cube [12].

In this manuscript, MD computer simulations of the deformation and subsequent plastic recovery of Ag and Cu layers with incoherent twin and cube-on-cube interfaces are compared with the microstructure in samples that exhibited plastic strain recovery; the samples studied were provided by Kingstedt *et al.* and the pertinent results are documented in Fig. 1 and Table 1 [12]. It will be shown that the essential features of the deformation processes observed experimentally are captured fully by the MD simulations. Specifically, and importantly, a rotation of the incoherent twin interface was observed experimentally and in the simulations despite the differences in test conditions. The capturing of this unique signature of the deformation response implies that similar mechanisms are active in the simulations and the experiments. Consequently, it is possible to use the simulations to reveal characteristics of the plastic strain recovery that are not accessible experimentally. The dynamics of plastic strain recovery were found to involve dislocation interactions with and accommodation within the interface, which is driven by the stress buildup at and a reduction of the energy of the interface. This mechanism will be shown to explain the dependence of the plastic strain recovery on the interface type as well as the loading direction with respect to the interface habit plane. Dislocation-dislocation interactions result in the formation of tangled structures or cell walls and, consequently, these dislocations cannot participate in supporting further strain. Similarly, the trapping of the dislocations will prohibit them from participating in the plastic strain recovery. It will be shown that dislocation tangles and even the formation of cell walls are more prevalent with increasing strain, which explains the dependence of the magnitude of the plastic strain recovery on the strain.

## 2. Methods

*2.1 Experiments*

The material used for this study was a directionally solidified AgCu eutectic alloy, which was processed to achieve a bilayer thickness of 500 nm with over 66 % of the interfaces being of the incoherent twin type and the remainder being of the cube-on-cube type [7]. The samples were subjected to either split Hopkinson pressure bar compression at a strain rate of $10^3$ s$^{-1}$ or quasi-static compression at a strain rate of $10^{-3}$ s$^{-1}$ using a load frame (the mechanical response of the compression tests are reported elsewhere [12, 17]); the key findings from these compression tests are summarized in Table 1. In Table 1, samples designated 1, 2 and 6 were prepared from one directionally solidified rod and samples 3, 4, and 5 were prepared from another; this was done to ensure the effect was not unique to one rod. To achieve high levels of strain and to follow the evolution of the recovery of plastic strain as a function of strain one sample was loaded, unloaded and recovered and then reloaded repeatedly.

Electron microscopy analysis was conducted on a Tecnai TF-30 TEM operating at 300 keV. Samples for TEM analysis were produced by using either a focused ion beam machining technique using a Zeiss Auriga with a final milling voltage of 2.0 keV or conventional ion milling preparation of 3 mm disks using a Fischione 1050 ion mill with a final milling voltage of 1.0 keV. With both TEM sample preparation techniques, the sample normal was near the growth axis of the AgCu rod.

*2.2 Molecular Dynamics Simulations*

MD simulations were performed using the LAMMPS software [18] and embedded atom force-field was employed to describe Ag-Ag, Ag-Cu, and Cu-Cu interactions [19]. The dislocation density was calculated using a dislocation extraction algorithm (DXA) [20]. All visualizations were conducted using OVITO [21]. The system configurations, shown in Fig. 2, mimic the multilayer structure of the Ag/Cu alloy used in the experiments. Periodic boundary conditions were applied along the x-, y- and z-directions, as defined in Fig. 2. Two Ag and two Cu layers alternate along the y-direction and are infinite along the x- and z-directions.

Two types of interfaces were considered in the simulations, incoherent twin interfaces with $(\bar{3}13)_{Ag}||(\bar{1}12)_{Cu}$ habit planes and cube-on-cube interfaces with $(\bar{1}11)_{Ag}||(\bar{1}11)_{Cu}$ habit planes. For AgCu alloys with incoherent twin interfaces, the Ag and Cu layers have different crystallographic orientations, as shown in Fig. 2(a). The dimensions of each of the Ag and Cu

layers were 38 nm in the x-direction and 35 nm in the z-direction. In the y-direction the Ag and Cu layers were 41 nm and 20 nm thick, respectively. For AgCu alloys with cube-on-cube interfaces, the Ag and Cu layers, as shown in Fig. 2(b), have the same crystallographic orientation. Each of the Ag and Cu layers were 36 nm in the x-direction and 35 nm in the z-direction. The Ag and Cu layers were 40 nm and 20 nm thick in the y-direction, respectively. Although these layer thicknesses are an order of magnitude smaller than the material used in the experiments, the 2:1 ratio of the thickness of Ag to Cu is consistent.

In order to construct the nanolayered structures, first single crystal Ag and Cu systems were generated. These single crystal systems were equilibrated for 1 ns at 5 K and 0 GPa using the Berendsen thermostat and barostat. Equilibrated Ag and Cu layers were then joined to form nanolayers stacked along the y-direction. Each sample had four Ag/Cu interfaces and consisted of over 10 million atoms. The energy of the as-stacked Ag/Cu interfaces was minimized using the conjugate gradient algorithm and then relaxed at 300 K and 0 GPa for another 1 ns.

For both systems, the compression tests were conducted by applying strain uniaxially along the z-direction at a strain rate of $5 \times 10^8$ s$^{-1}$. The strain was applied by alternately deforming the system by 0.05 % and equilibrating it for 1 ps. During compression, the Nose-Hoover thermostat was used in the x-, y- and z-directions to maintain the temperature at 300 K and the Nose-Hoover barostat was applied to the x- and y-directions to maintain zero stresses and to allow relaxation of the system along those directions. The maximum applied compressive strains were varied between 6 % and 25 %, which is consistent with the range of strains used in the experiments. Recovery was simulated by removing the constraint on the z-dimension of the sample and by performing simulations at a constant temperature and pressure, where the pressure was set to 0 GPa.

## 3. Results

*3.1 Initial structures*

The microstructure of the directionally solidified material is predominantly lamellar (>66 %), although regions of globular platelets of Cu in a Ag matrix exist (<34 %) [7]. These structures are elongated in and are discontinuous along the growth direction. The layer thicknesses of the phases are approximately 170 nm and 330 nm for Cu and Ag, respectively. The lamellar morphology consists of an incoherent twin interface structure between Ag and Cu, as shown in Figs. 3(a)-(c). The interiors of both the Ag and Cu layers are mostly dislocation free. The interfaces, however, contain a higher density of dislocations, which are present as half loops terminating at the interfaces; examples are marked by arrowheads in the bright-field TEM micrograph presented in Fig. 3(a). From the selected area diffraction pattern, shown in Fig. 3(a), the orientation relationship between the phases is approximately 2° away from that of a twin about the $[101]_{Ag}||[110]_{Cu}$ growth axis; this deviation is evident from comparison of the $(11\bar{1})_{Ag}$ and $(11\bar{1})_{Cu}$ diffraction spots, which are not collinear. The interface habit planes are incoherent twin and predominantly $(\bar{3}13)_{Ag}||(\bar{1}12)_{Cu}$ with variations toward the coherent twin plane of up to 10°, but more commonly up to 5°. The detailed structure of the incoherent twin interfaces is shown in the high-resolution TEM (HRTEM) micrograph presented in Fig. 3(b). The interfaces are stepped, with $(\bar{1}11)_{Ag}$ and $(\bar{1}\bar{1}1)_{Ag}$ planes on the Ag side and $(\bar{1}11)_{Cu}$ and $(002)_{Cu}$ planes on the Cu side. MD simulations of the Ag-Cu multilayers with incoherent twin interfaces, Fig. 3(c), show the same arrangement of atoms found in the HRTEM micrograph.

The second type of Ag/Cu interface that exists in this material has a cube-on-cube orientation relationship with interface habit planes of either $(\bar{1}11)_{Ag,Cu}$ or $(11\bar{1})_{Ag,Cu}$. With this interface, all of the crystallographic planes and directions between Ag and Cu are aligned. A region of this interface type is shown in the bright-field micrograph and selected area diffraction pattern presented in Fig. 3(d). Dislocations present in the initial material are primarily half loops in the Ag layer terminating at the interface. Examples of such dislocation loops are marked by arrowheads in Fig. 3(d). The HRTEM micrograph in Fig. 3(e) shows that the cube-on-cube interface is atomically flat with randomly spaced steps. Curved sections of the interfaces are a result of alternating segments of the $(\bar{1}11)_{Ag,Cu}$ and $(11\bar{1})_{Ag,Cu}$ planes. Misfit dislocations are observed at periodic intervals of 8-11 atomic planes of Cu and examples of these are marked by circles in Fig. 3(e). In addition to the periodic dislocation structure, there is evidence of elastic strain at the interface, examples are marked by arrows in Fig. 3(e). The MD simulation for Ag-Cu multilayers with this orientation relationship, Fig. 3(f), shows nearly the same atomic-level structure as the interface observed experimentally. Specifically, the simulated structures contain an atomically flat $(\bar{1}11)_{Ag,Cu}$ interface with periodic misfit dislocations every nine Cu $(11\bar{1})_{Cu}$ atomic layers.

*3.2 Mechanical response of incoherent twin interfaces*

Stress-strain curves from MD simulations after compressing the material with incoherent twin interfaces along $[101]_{Ag}||[110]_{Cu}$ to strains of 10 % and 25 % are shown in Fig. 4(a). The curves have a distinct upper and lower yield point, which is followed by a region that shows a low work hardening rate. In this regard the simulations do not yield stress-strain curves consistent with those observed experimentally, compare Fig. 4 with Fig. 1 [12, 17]. Dislocation nucleation for both 10 % and 25 % strain simulations occurred at the maximum stress, which corresponded to an elastic strain of 3.8 %. After the systems were compressed uniaxially along the z-axis to the desired strain, the external constraint on the z-dimension was removed and the recovery process began. In the first stage of recovery, 1 ps after unloading, the strain decreased from an initial value of 10.00 % to 9.08 % as the average stress decreased to 0 GPa. This change in strain was due mainly to elastic recovery. Over the next 500 ps, the strain continued to decrease gradually to 8.25 %, which was due to plastic recovery that will be shown to be due to internal stresses. The change in strain with time during recovery is shown in the inset of Fig. 4(a). To estimate the error bar in the measured plastic strain recovery, we have performed eight simulations, all for the case of 10 % strain but with different initiation conditions. The error bar corresponding to 97.5 % confidence interval calculated using Student's t-distribution is 0.047 %, which is 1-2 orders of magnitude smaller than the average values of plastic strain recovery observed in our simulations. Higher strains were also simulated and it was found that as strain increased the amount of plastic recovery decreased. The degree of plastic recovery, both in the experiments and in the simulations, as a function of compressive strain is shown in Fig. 4(b), and is summarized in Tables 1 and 2, respectively. In comparison to the experiments, the magnitude of the measured plastic recovery is lower in the simulations. Potential reasons for this difference include: the time available for plastic recovery in the experiments was on the order of minutes whereas in the simulations it was on the order of nanoseconds and the density of the dislocations are likely different in the simulations and experiments. The latter potential difference could not be verified as the dislocation density prior to the plastic recovery is not accessible experimentally. However, both experiments and simulations show plastic strain recovery and that the magnitude of the recovery decreases with increasing strain. The DXA algorithm was used to quantify the dislocation density in the simulations after 10 % strain and then after 500 ps recovery. The results, presented in Fig. 4(c), show that during recovery the dislocation density decreased from $9.8\times10^{12}\text{cm}^{-2}$ to $7.3\times10^{12}\text{cm}^{-2}$ and from $3.8\times10^{12}\text{cm}^{-2}$ to $2.9\times10^{12}\text{cm}^{-2}$ in the Ag and Cu layers, respectively. Unfortunately, this change in the dislocation density cannot be confirmed experimentally.

A comparison of the dislocation density in the Ag and Cu layers in samples strained dynamically that exhibited plastic strain recovery (sample 1, 3.2% plastic strain recovery) and ones that did not (sample 2, 4$^{\text{th}}$ loading) are compared in Fig. 5. For completeness, the microstructure developed under quasi-static loading, which also showed no recovery, is included in Fig. 5. In comparing the dislocation structures it is evident that the dislocation density is always highest in the Ag layers, and that in samples that exhibited no plastic strain recovery the dislocations in the Ag were tangled (quasi-static, Fig. 5(c)) or organized into dislocation cells (sample 2, 4$^{\text{th}}$ loading, Fig. 5(b)). The primary difference in the Cu layer was that the dislocation density was lowest in the sample that exhibited plastic strain recovery. The locking of dislocations with each other provides a possible explanation for the decrease in plastic recovery with increasing strain.

The decrease in the dislocation density on recovery cannot be determined experimentally, however, a unique signature of this deformation response was that it caused a change in the interface character corresponding to an additional rotation away from the twin orientation relationship. The magnitude of this rotation and the loading conditions under which it occurred are summarized in Table 3. Representative zone-axis TEM micrographs and corresponding selected area diffraction patterns of samples loaded dynamically to strains of 9.3 % and 21.9 % and quasi-statically to a strain of 4 % are compared in Fig. 6. The rotation is measured by the angle between the $[11\bar{1}]_{\text{Ag}}$ and $[11\bar{1}]_{\text{Cu}}$ diffraction spots, and is visualized by the dotted lines in the selected area diffraction patterns in Figs. 6(d)-(f). Rotation away from the twin orientation occurred for all loading conditions involving incoherent twin interfaces loaded along the $[101]_{Ag}||[110]_{Cu}$ growth direction. For loading dynamically along $[101]_{Ag}||[110]_{Cu}$, the rotation away from the twin orientation relationship increases with increasing strain, as shown in Table 3. Furthermore, even loading quasi-statically to a strain of 4.0 %, the interface rotated although no plastic recovery was observed. Interestingly, the degree of rotation observed for the quasi-static loading was similar in magnitude to the sample loaded dynamically to a strain of 9.3 %. However, it is worth noting that the latter sample exhibited recovery and the final strain was 5.7 %.

Analogous to the measurements taken experimentally, the angle between the $(\bar{1}11)$ plane in both Ag and Cu phases was measured from the Fast Fourier transforms of the undeformed, deformed, and deformed and recovered simulated layer structures. The simulated atomic structures are shown in Figs. 7(a)-(c) and the corresponding Fourier transforms can be found in Figs. 7(d)-(f). The rotation is measured as the angle between $[\bar{1}11]_{Ag}$ and $[\bar{1}11]_{Cu}$ and is indicated by the dotted lines in Figs. 7(d)-(f). The values of this angle for several of the simulations are summarized in Table 3. At peak loading the angle was found to be 6.6° at 10 % strain, and after 1 ns of recovery it decreased to 6.0°. This decrease did not occur after recovery following straining to 25 %, indicating that the mechanism which is responsible for the recovery and the decrease in the angle after loading to 10 % strain is not active at the higher strain.

Experimentally, loading the incoherent twin interfaces 90° to the growth direction to 15.4 % strain did not result in a change in orientation relationship, that is, the angle between the $[\bar{1}11]_{Ag}$ and $[\bar{1}11]_{Cu}$ did not change. The deformed microstructures contained dislocations primarily in the Ag phase and these dislocations were organized into a dislocation cell structure.

Deformation twinning was also observed to a limited extent at this load orientation and was confined to the Ag phase [7]. A representative bright-field micrograph of the deformation microstructure after loading 90° to the growth direction is shown in Fig. 8(a) and a corresponding selected area diffraction pattern in Fig. 8(b). An analogous MD simulation was performed, the system with incoherent twin interfaces, Fig. 2(a), was compressed along $[\bar{3}13]_{Ag}$ and $[\bar{1}12]_{Cu}$, which corresponds to loading 90° to the growth direction in the experiments. Numerical Fourier transformations conducted on atomic positions of the multilayers observed from the $[101]_{Ag}$ and $[110]_{Cu}$ direction show that the angle between $[\bar{1}\,1\,1]_{Ag}$ and $[\bar{1}\,1\,1]_{Cu}$ is approximately 2.6° before compression and does not change after straining to 10 % or after 1 ns of recovery.

Experiments and simulations of the cube-on-cube interface type reveal the orientation relationship between Ag and Cu remains cube-on-cube after deformation. The on-zone bright-field TEM image presented in Fig. 8(c) shows the deformation microstructure in a region with cube-on-cube interfaces after 9.5 % strain. Dislocations are found mainly in the Ag phase, with a higher density in regions of curvature of the Cu phase. The corresponding selected area diffraction pattern in Fig. 8(d) shows the orientation relationship between the two phases remains unchanged. Simulations also showed a retention of the orientation relationship between the two phases, because, unlike the incoherent twin interface case, the shape change of the layers due to dislocation slip was the same for both the Ag and Cu layers.

Molecular dynamics simulations can provide details of dislocation activity during recovery in either the Ag or the Cu layer. The interactions were similar in nature in both layers and for ease of describing different dislocation processes, representative examples from the Cu layer are presented. Figure 9 shows dislocation activity in layers with incoherent twin interfaces and illustrates the mechanisms of plastic recovery. During compression, dislocations were nucleated at the interfaces and these were observed to glide into interior regions of the layers. The observation of the interfaces serving as dislocation sources is consistent with the experimental observations of dislocation half-loops emerging from the interfaces. Different dislocation slip systems were activated and as these involved partial dislocations, the interaction of them resulted in the formation of stair-rod dislocations. As compression continued, dislocations were pinned either at the interfaces, or by other defects including other dislocations and stacking-fault tetrahedra. During recovery, the dislocation structure changed. Some dislocations interacted with other isolated dislocations or complex dislocation structures to become part of the immobile dislocation network. Other dislocations interacted with the interfaces and were incorporated into them. To illustrate two of the retraction mechanisms, activities of two representative interactions are presented in Fig. 9 and discussed in detail below. Again, for ease of discussion of the plastic recovery mechanism, the examples are selected from a Cu layer but similar interactions were observed to occur in the Ag layers.

The first example shows glissile partial dislocations generated from a dislocation tangle interacting with an incoherent twin interface and with other complex dislocation tangles in the Cu layer during loading and the reverse motion of these partial dislocations and their eventual incorporation back into the dislocation tangle from which they were generated on unloading. The Burgers vectors of the leading and trailing partial dislocations are $b_{\text{lead}} = \frac{a}{6}[11\bar{2}]$ and $b_{\text{trail}} = \frac{a}{6}[2\bar{1}\bar{1}]$, respectively. This pair of dislocations was generated from a stair-rod dislocation as part of the evolution of one of the complex dislocation tangles that exists in the Cu layer; the nucleation event is not captured in the images presented in Fig. 9, but the dislocation tangle from which they were generated is labeled E. The dislocation tangle results from the intersection of partial dislocations on different {111} planes, which yields a stair rod dislocation at the intersection; another example of such a complex dislocation tangle is labeled D in Fig. 9. The glissile partial dislocations, along with the stacking fault (yellow atoms), following loading to a strain of 9.95 % are shown in Fig 9(a) to be bowed out between pinning points at locations A and B. The pinning points are dislocation debris. The direction of motion of the leading partial dislocation is indicated by arrows in Fig. 9(a). The leading partial dislocation is blocked by the dislocation structure D. The intersection is with a partial dislocation with a Burgers vector of $\frac{a}{6}[\bar{1}\bar{2}1]$, which resides on $(\bar{1}11)_{Cu}$, and intersects structure D beneath the viewing plane. This interaction resulted in the formation of a stair-rod dislocation, $\frac{a}{6}[\bar{1}\bar{2}1] + \frac{a}{6}[11\bar{2}] = \frac{a}{6}[0\bar{1}\bar{1}]$. As the strain was increased, the leading and trailing partial dislocations bowed out further between the pinning points and the leading partial dislocation intersected the interface and was blocked at location C, Fig. 9(b). On removing the external load, the leading partial dislocation was released from the interface pinning point and started to retract. After 9.5 ps from the time the external stress was removed, the two partial dislocations had retracted partially as shown in Fig. 9(c) but the reverse motion of the leading partial dislocation was hindered by the stair-rod dislocations generated from the intersection with defect structure D. After 15.5 ps, the leading partial dislocation was released from structure D and the stair-rod dislocation was eliminated. The leading and trailing partial dislocations remained pinned between pinning points A and B and started to bow out in the opposite direction, Fig. 9(d). With the release of the leading partial dislocation from structure D, it started to retract; this retraction can be seen by comparing Figs 9(c) and 9(d). On release of both partial dislocations from pinning point A, they move toward defect structure E, as shown in Fig. 9(e). At the recovery time of 27 ps, what was initially the trailing partial dislocation interacts with the dislocations within the complex structure E, which essentially results in assimilation of these partial dislocations into structure E, Fig. 9(f). This series of images illustrates the role of the dislocation structure in the evolution of the microstructure during straining and during recovery.

The second example shows the interaction and annihilation of a layer of stacking fault through the interaction of a partial dislocation with the interface. The example shown in Figs. 9(g)-(i) is within the Cu layer. The leading partial dislocation, marked by the arrow in Fig. 9(g), lies on the $(111)_{Cu}$ plane and has a Burgers vector of $\frac{a}{6}[\bar{1}2\bar{1}]$ and is pinned at the interface at the locations marked by arrowheads. On removing the applied load, it retracts back to and spreads along the interface as seen in Figs. 9(h) and (i). Between Figs. 9(g) and 9(h), the dislocation retracts towards the interface and extends along it toward the dislocation structure F, which resides on $(11\bar{1})_{Cu}$. At a recovery time of 50 ps, this partial dislocation had retracted to and spread along the interface in addition to part of the dislocation combining with structure F. The dislocation spreading along the interface and combining with dislocation content in the interface will cause rearrangement within the interface. There is also a reduction in the faulted area.

The change in interfacial structure and build-up of dislocations at the interfaces can be further interpreted from the perspective of the potential energy. Plotting local potential energy as a function of the distance from the interface provides a practical way to access information about the effect of the loading history on the evolution of interfacial energy. This is because interfacial energy is equal to the difference in potential energies of the interface and a corresponding single crystal, normalized by the interfacial area. The local potential energy is plotted in Figs. 10(a)-(c) for the cases of (a) incoherent twin interfaces with the load applied along the $[101]_{Ag}$ and $[110]_{Cu}$ directions, (b) incoherent twin interfaces with the load applied 90° away from the experimental growth direction, specifically along $[\bar{3}13]_{Ag}$ and $[\bar{1}12]_{Cu}$, and (c) cube-on-cube interfaces with loading along the $[101]_{Ag,Cu}$. The reported potential energy is averaged over 4 Å-thick slices parallel to the Ag/Cu interfaces (x-z plane). For the multilayer simulation with incoherent twin interfaces before compression (0 % strain), the energy at the interface region is slightly higher (by 0.0024 eV) than the energy in the Ag layer. After the samples are compressed to a strain of 10 % with the loading direction along the $[101]_{Ag}$ and $[110]_{Cu}$ direction, the interfacial energy increased from -2.8096 eV to -2.7968 eV and the energy of the inner Ag layers increased from -2.8120 eV to -2.8082 eV. After recovery, the interfacial energy decreased to -2.8000 eV and the energy of the interior of the Ag layers decreased to -2.8091 eV. The decrease in energy at the interfaces (by 0.11 %) is more significant than that inside the layers (by 0.03 %), Fig. 10(a).

After loading the same multilayer system with incoherent twin interfaces to a compressive strain of 10 %, but this time with the loading direction 90° away from $[101]_{Ag}$ and $[110]_{Cu}$, specifically, along $[\bar{3}13]_{Ag}$ and $[\bar{1}12]_{Cu}$, the interfacial energy increased from -2.8096 eV to -2.7976 eV and the energy of interior of the Ag layers increased from -2.8120 eV to -2.8096 eV, Fig. 10(b). After recovery, the interfacial energy and the energy of the interior of Ag layers decreased to -2.7986 eV and -2.8104 eV, respectively. A decrease in energy at the interfaces (0.04 % decrease) is also greater than the decrease of energy inside layers (0.03 % decrease), though the difference is not as significant as in the case when the samples were loaded along $[101]_{Ag}$ and $[110]_{Cu}$.

A similar analysis for the multilayered structure with cube-on-cube interfaces loaded along $[101]_{Ag,Cu}$ was performed. Here, before compression (0 % strain), the average energy in the interface region and the average energy inside the Ag layers was the same and equal to -2.8122 eV. The energy is distributed more evenly among the interfaces and the layer interiors than in the case of incoherent twin interfaces. After the samples with the cube-on-cube interfaces were compressed to a strain of 10 %, the interfacial energy increased to -2.7986 eV and the energy of the interior of the Ag layers increased to -2.8021 eV. After recovery, the interfacial energy and the energy of the interior of the Ag layers decreased to -2.8035 eV and -2.8051 eV, respectively (see Fig. 10(c)). Again, the energy decrease at the interfaces (by 0.18 %) is greater than in the layer interiors (0.11 %), but the difference between the layer interior and the interface is not as significant as that observed in the multilayers with incoherent twin interfaces loaded along the $[101]_{Ag}$ and $[110]_{Cu}$ direction.

Figure 10(d) shows the ratio of energy change at the interfaces to the energy change in the Ag layer interiors after 500 ps recovery. The ratios are shown for the multilayers with incoherent twin interfaces loaded along the $[101]_{Ag}$ and $[110]_{Cu}$ direction, and loaded along $[\bar{3}13]_{Ag}$ and $[\bar{1}12]_{Cu}$, as well as for the multilayers with cube-on-cube interfaces loaded along $[101]_{Ag,Cu}$. In all three cases the ratio is larger than 1, which means that the energy change in the interfacial region is more significant than that in the Ag layer interiors. There are two reasons for this ratio being larger than 1. The first is that during deformation, dislocation content builds up at the interfaces to a larger extent than the layer interiors. As a result, during recovery the dislocation density near the interfaces decreases more significantly than the dislocation density in the Ag layer interiors. The second is that dislocations blocked by the interfaces can be repelled from the interfaces by the local stress after the external load is removed.

**4. Discussion**

Comparison of the experimental and simulation results provides insights into the plastic strain recovery mechanisms, which are based on reversing the direction of dislocation motion, dislocation annihilation, and rearrangements at and within the interfaces. These mechanisms are dependent on the activated slip systems resulting in a change in interface structure and increasing the energy of the interfaces, providing the driving force for the plastic recovery. Specifically, the change of the interface structure, in terms of orientation relationship , is proposed to be the driving force behind the plastic recovery. The degree of this recovery was dependent on the loading direction and the interface type; this was evident in the experiments but not the simulations. The degree of plastic strain recovery was greatest for loading along the $[101]_{Ag}$ and $[110]_{Cu}$ growth direction when the interfaces were incoherent twins. Specific to loading in this direction, the simulations showed that preferential deformation in the $[0\bar{1}0]_{Ag}$ and $[00\bar{1}]_{Cu}$ direction occurs and this causes the change in the orientation relationship between Ag and Cu. When the original orientation relationship between the Ag and Cu layers is cube-on-cube, and both phases are loaded along the $[101]_{Ag,Cu}$ direction, the layers still deform anisotropically but in the same direction for Ag and Cu, and as a consequence retain their original orientation relationship. Similarly, loading of coherent twin interfaces 90° to the growth direction, along $[\bar{3}13]_{Ag}$ and $[\bar{1}12]_{Cu}$, did not change the orientation relationship between the layers. For this case the magnitude of the plastic recovery was similar in the simulations and experiments and was less than loading along the $[101]_{Ag}$ and $[110]_{Cu}$ growth direction.

The dependence of the magnitude of the recovery on the interface type and loading direction can be made based on the ratio of energy change between the interface and layer interiors, see Fig. 10(d). This ratio is larger for the case which shows the largest plastic strain recovery at low levels of strain. That is, it is larger for loading of incoherent twin interfaces along $[101]_{Ag}$ and $[110]_{Cu}$, Fig. 10(a). The difference in this ratio is related to the slip systems that become active during compression in each loading geometry. For example, loading along the $[101]_{Ag}$ and $[110]_{Cu}$ directions for the incoherent twin interfaces activates dislocations that are blocked by the interfaces and also causes a change in orientation relationship between the Ag and Cu layers. In contrast, loading 90° from the $[101]_{Ag}$ and $[110]_{Cu}$ directions, when the interfaces are incoherent twins, activates some dislocations that can transfer across the interfaces and, additionally, the orientation relationship between the Ag and Cu layers is not changed by the deformation. Similarly, for cube-on-cube type interfaces loaded along the $[101]_{Ag,Cu}$ axis, the original orientation relationship is retained and some specific dislocations can transfer readily across the interfaces. The retention of the initial orientation relationship in the latter two cases results in a smaller local residual stress at the interface. This means incoherent twin interfaces with the applied load along $[101]_{Ag}$ and $[110]_{Cu}$ are more effective in decreasing dislocation density in the interface region after unloading, leading to plastic strain recovery, and resulting in a larger ratio of energy change between peak loading and recovery. While similar to the back motion of dislocations for plastic strain recovery cited in the penta-twinned Ag nano-wires [15], the AgCu eutectic, however, appears to require a change in the interface structure to drive the plastic strain recovery.

To explore the decrease in magnitude of the plastic strain recovery with increasing strain and the lack of any significant plastic strain recovery for quasi-statically deformed specimens with primarily incoherent twin interfaces compressed along $[101]_{Ag}$ and $[110]_{Cu}$, it is necessary to consider the evolved microstructure within the layers. In both cases, the evolved deformation microstructures consist of dislocation cell structures and dislocation tangles, Fig. 5. The lack of plastic recovery can be attributed to the dislocation-dislocation interactions locking the dislocations in position, which inhibits dislocation motion and, hence, the plastic strain recovery.

5. Conclusion

The mechanism for plastic strain recovery in AgCu eutectic was determined to be dislocation retraction and annihilation, which is driven by stresses at the interfaces. This was most prominent for the case in which plastic deformation in the Ag and Cu layers was incompatible, such that the orientation relationship between Ag and Cu changes during loading. Dislocations on the activated glide systems were also found to be blocked, and residual stresses built-up at the interfaces promoting dislocation back motion during unloading. In the cases in which the deformation in the Ag and Cu layers were compatible and there is no change in the orientation relationship, plastic strain recovery was inhibited. In addition, only some of the activated dislocations were blocked by the interfaces. As a result, the driving force for dislocation back motion exerted by these interfaces was smaller than in the case with plastic strain recovery. Decreasing plastic strain recovery with increasing strain as well as lack of plastic strain recovery in quasi-statically loaded specimens is attributed to the locking of dislocations in dislocation tangles and cell structures in the layer interiors.

**Acknowledgments**

BPE and IMR acknowledge the contributions of Dr. O. Kingstedt and Prof. J. Lambros for providing the deformed samples and for many fruitful discussions. The experimental work was performed, in part, at the University of Wisconsin Madison and was supported by the US Department of Energy Office of Basic Energy Sciences, Division of Materials Science, under award No. DEFG-02-07ER46443 (IMR and BPE). Simulation work was supported by the Army Research Office grant # W911NF-12-1-0548. Instrument support was also provided by Materials Research Science and Engineering Center (DMR-1121288) and Nanoscale Science and Engineering Center (DMR-0832760) at University of Wisconsin-Madison.

**References**

[1]  A. Balyanov, J. Kutnyakova, N.A. Amirkhanova, V.V. Stolyarov, R.Z. Valiev, X.Z. Liao, Y.H. Zhao, Y.B. Jiang, H.F. Xu, T.C. Lowe, Y.T. Zhu. Corrosion resistance of ultra fine-grained Ti, Scripta Materialia 51 (2004) 225-229.

[2]  P.Q. La, J.Q. Ma, Y.T. Zhu, J. Yang, W.M. Lu, Q.J. Xue, R.Z. Valiev. Dry-sliding tribological properties of ultrafine-grained Ti prepared by severe plastic deformation, Acta Materialia 53 (2005) 5167-5173.

[3]  J.T. Wood, J.D. Embury, M.F. Ashby. An approach to materials processing and selection for high-field magnet design, Acta Materialia 45 (1997) 1099-1104.

[4]  T. Hochbauer, A. Misra, K. Hattar, R.G. Hoagland. Influence of interfaces on the storage of ion-implanted He in multilayered metallic composites, Journal of Applied Physics 98 (2005) 1-7.

[5]  K. Hattar, M.J. Demkowicz, A. Misra, I.M. Robertson, R.G. Hoagland. Arrest of He bubble growth in Cu-Nb multilayer nanocomposites, Scripta Materialia 58 (2008) 541-544.

[6]  M.A. Meyers, A. Mishra, D.J. Benson. Mechanical properties of nanocrystalline materials, Prog. Mater. Sci. 51 (2006) 427-556.

[7] B.P. Eftink. Dislocation interactions with characteristic interfaces in Ag-Cu eutectic. Materials Science and Engineering, vol. Ph.D.: University of Illinois Urbana-Champaign, 2016.

[8] S.J. Zheng, J. Wang, J.S. Carpenter, W.M. Mook, P.O. Dickerson, N.A. Mara, I.J. Beyerlein. Plastic instability mechanisms in bimetallic nanolayered composites, Acta Materialia 79 (2014) 282-291.

[9] Y.Z. Tian, Z.F. Zhang. Bulk eutectic Cu-Ag alloys with abundant twin boundaries, Scripta Materialia 66 (2012) 65-68.

[10] O.T. Kingstedt, B. Eftink, J. Lambros, I.M. Robertson. Quasi-static and dynamic compressive deformation of a bulk nanolayered Ag–Cu eutectic alloy: Macroscopic response and dominant deformation mechanisms, Materials Science and Engineering: A 595 (2014) 54-63.

[11] J. Kacher, B.P. Eftink, B. Cui, I.M. Robertson. Dislocation interactions with grain boundaries, Curr. Opin. Solid State Mat. Sci. 18 (2014) 227-243.

[12] O.T. Kingstedt, B.P. Eftink, I.M. Robertson, J. Lambros. Inelastic strain recovery of a dynamically deformed unidirectional Ag-Cu eutectic alloy, Acta Materialia (accepted) (2016).

[13] J. Rajagopalan, J.H. Han, M.T.A. Saif. Plastic deformation recovery in freestanding nanocrystalline aluminum and gold thin films, Science 315 (2007) 1831-1834.

[14] X.Y. Li, Y.J. Wei, W. Yang, H.J. Gao. Competing grain-boundary- and dislocation-mediated mechanisms in plastic strain recovery in nanocrystalline aluminum, Proc. Natl. Acad. Sci. U. S. A. 106 (2009) 16108-16113.

[15] Q.Q. Qin, S. Yin, G.M. Cheng, X.Y. Li, T.H. Chang, G. Richter, Y. Zhu, H.J. Gao. Recoverable plasticity in penta-twinned metallic nanowires governed by dislocation nucleation and retraction, Nat. Commun. 6 (2015) 1-8.

[16] R.Z. Li, H.B. Chew. Deformation twinning and plastic recovery in Cu/Ag nanolayers under uniaxial tensile straining, Philos. Mag. Lett. 94 (2014) 260-268.

[17] O.T. Kingstedt, B.P. Eftink, I.M. Robertson, J. Lambros. Anisotropic dynamic compression response of a directionally-cast silver–copper eutectic alloy, Acta Materialia 105 (2016) 273-283.

[18]   S. Plimpton. Fast parallel algorithms for short-range molecular dynamics, J. Comput. Phys. 117 (1995) 1-19.

[19]   P.L. Williams, Y. Mishin, J.C. Hamilton. An embedded-atom potential for the Cu-Ag system, Model. Simul. Mater. Sci. Eng. 14 (2006) 817-833.

[20]   A. Stukowski, K. Albe. Dislocation detection algorithm for atomistic simulations, Model. Simul. Mater. Sci. Eng. 18 (2010) 1-15.

[21]   A. Stukowski. Visualization and analysis of atomistic simulation data with OVITO-the Open Visualization Tool, Model. Simul. Mater. Sci. Eng. 18 (2010) 1-7.

**Figure captions**

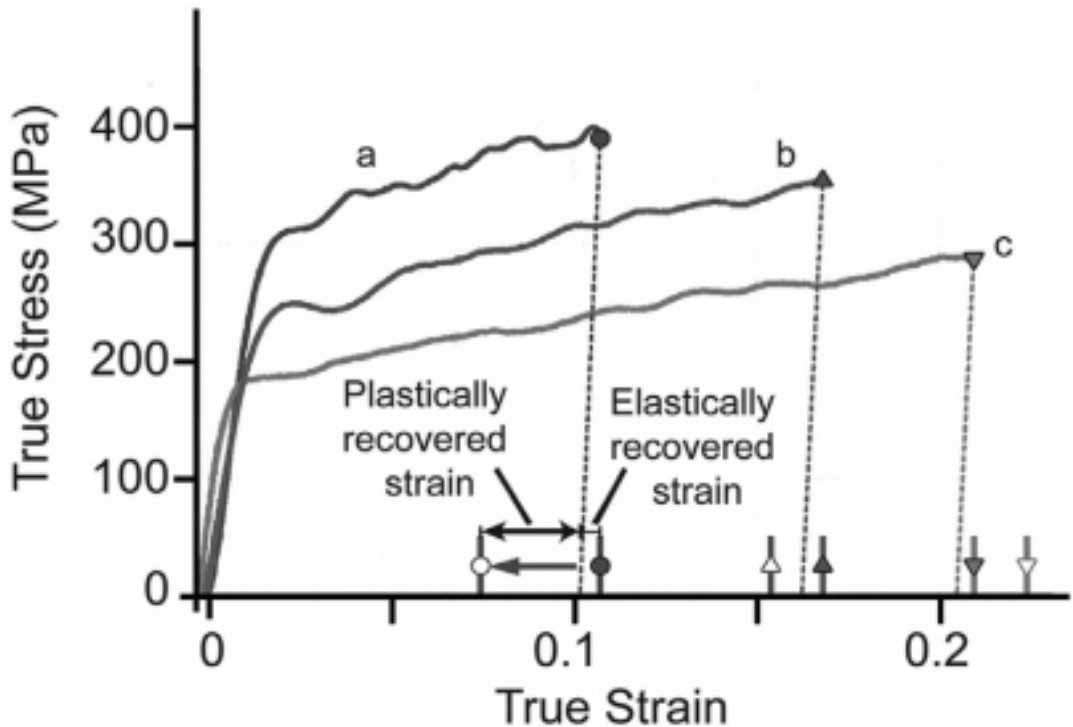

Figure 1: Stress-strain curves for directionally solidified AgCu eutectic loaded with split-Hopkinson pressure bar (SHPB) at: a 0°; b 90° and c 45° to the growth direction. Closed

symbols mark strain applied at a strain rate of $10^3$ s$^{-1}$, and open symbols mark strain measured from the recovered samples. Adapted from Kingstedt *et al.*[12].

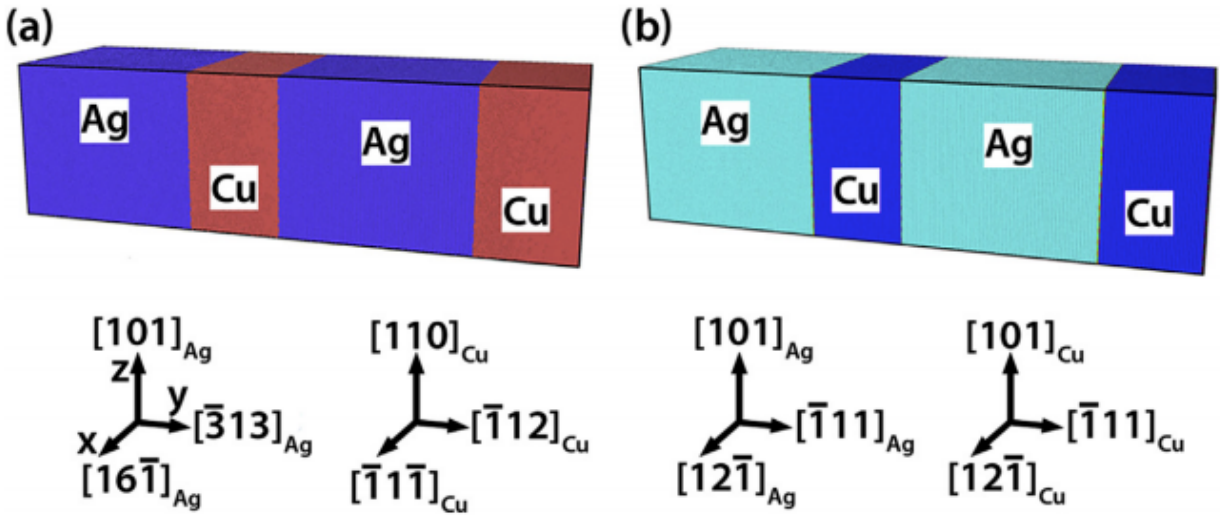

Figure 2: Molecular dynamics simulation setup for (a) Ag/Cu incoherent twin interfaces and (b) Ag/Cu cube-on-cube interfaces.

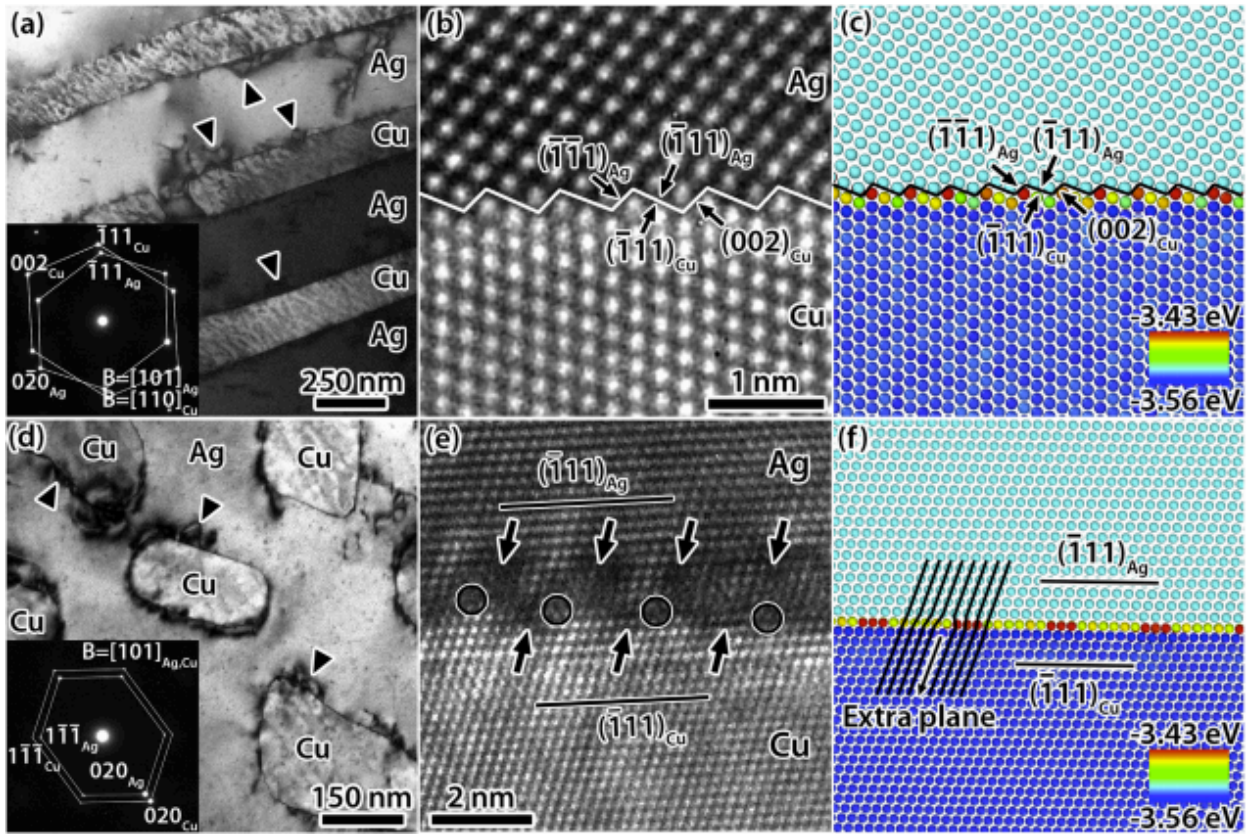

Figure 3: (a) Bright-field TEM micrograph of incoherent twin Ag/Cu interfaces with selected area diffraction pattern shown in the inset. (b) HRTEM micrograph of an incoherent twin Ag/Cu interface. (c) MD simulation of incoherent twin Ag/Cu interface. (d) Bright-field TEM micrograph of cube-on-cube Ag/Cu interfaces with selected area diffraction pattern in the inset. (e) HRTEM micrograph of a cube-on-cube Ag/Cu interface. (f) MD simulation of a cube-on-cube Ag/Cu interface. Arrowheads in (a) and (d) mark dislocations in Ag at Ag/Cu interfaces. Arrows and circles in (e) mark periodic elastic strain and misfit dislocations, respectively. In (c) and (f), Ag atoms are colored with cyan, and Cu atoms are colored based on calculated potential energy.

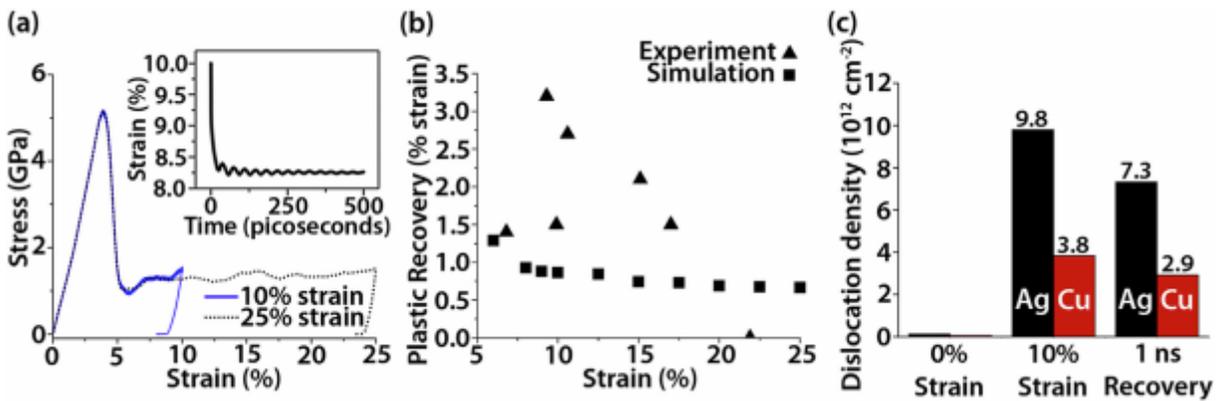

Figure 4: (a) Stress-strain relation during compression and recovery of the Ag-Cu system with incoherent twin interfaces from MD simulations. Inset shows the change in strain with time during recovery when strained to 10 %. (b) Recovered plastic strain as a function of compressive strain present in the samples before recovery. (c) Dislocation density in Ag and Cu layers measured in MD simulations.

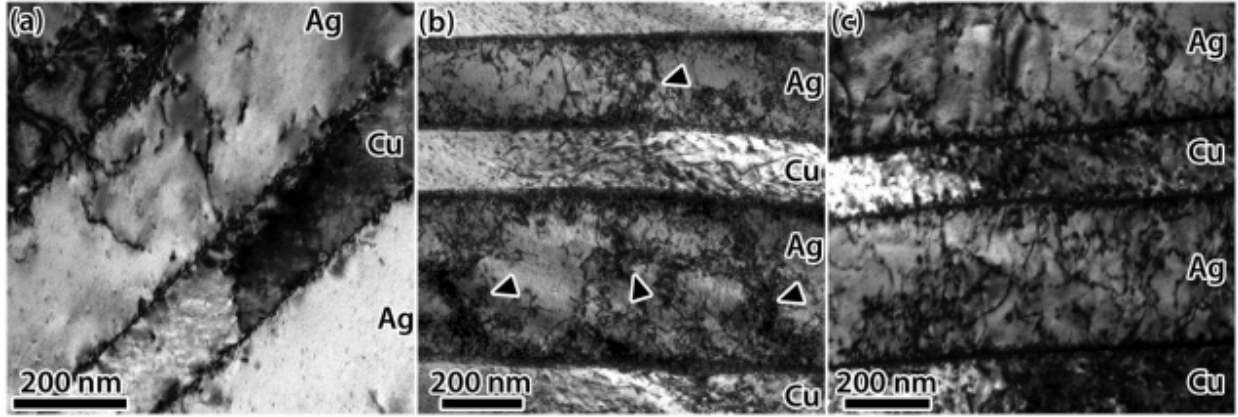

Figure 5: Bright-field TEM micrographs of regions with incoherent twin interfaces (a) after loading to a true strain of 9.3 % dynamically; (b) after loading dynamically to a true strain of 21.9 %; (c) after loading to a true strain of 4 % by quasi-static compression. (d), (e), and (f) are selected area diffraction patterns of (a), (b), and (c), respectively. Arrowheads in (b) mark dislocation cell walls.

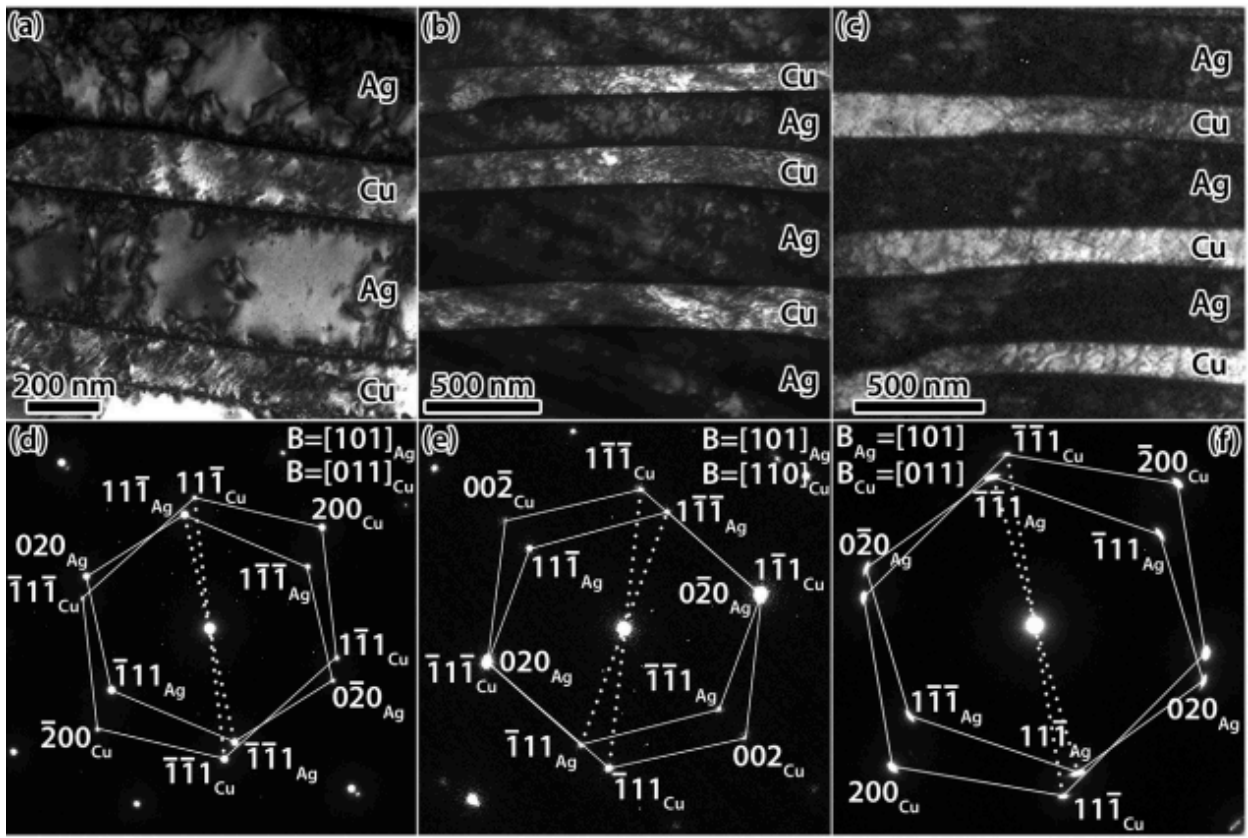

Figure 6: Bright-field TEM micrographs of regions with incoherent twin interfaces after loading to dynamic strains of (a) 9.3 %, and (b) 21.9 % and quasi-statically to a strain of (c) 4 %.

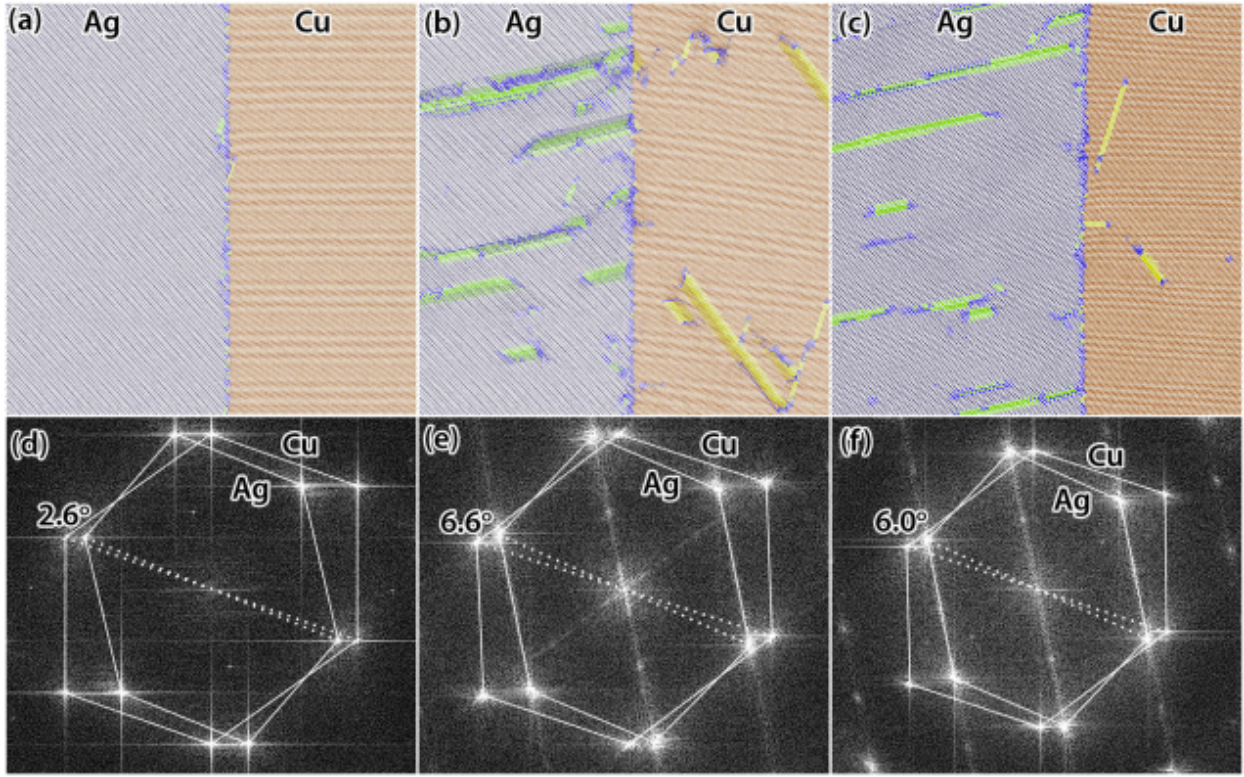

Figure 7: Simulated atomic structures containing incoherent twin interfaces visualized in real space (a) before compression, (b) after compression at the peak strain of 10 % and (c) after recovery. (d), (e) and (f) are Fourier transforms of (a), (b), and (c), respectively. The direction normal to the image plane is $[101]_{Ag}$ and $[110]_{Cu}$. The difference between the $(\bar{1}11)_{Ag}$ and $(\bar{1}11)_{Cu}$ spots is visualized by the white dotted lines in (d), (e), and (f).

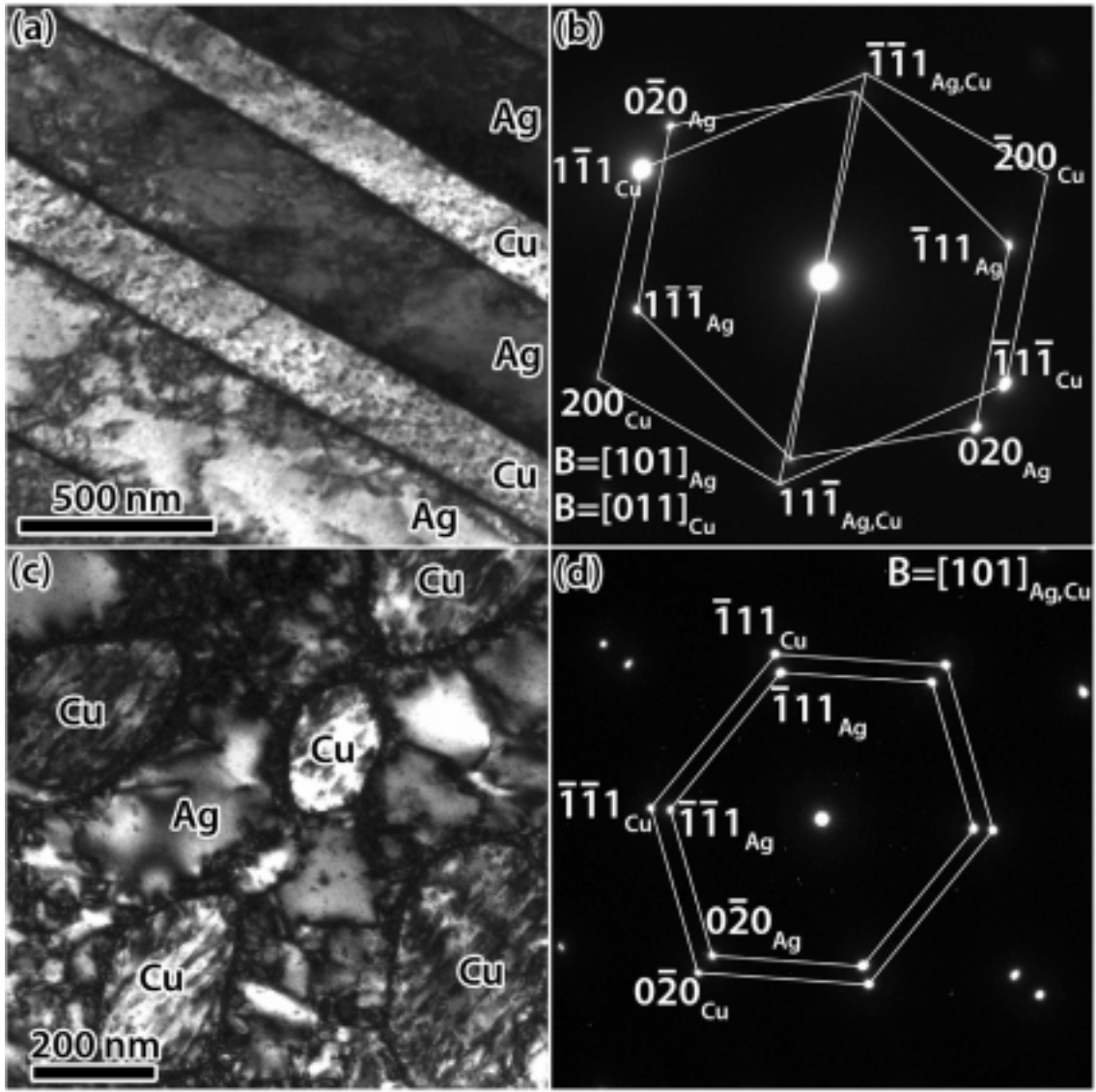

Figure 8: (a) Bright-field TEM micrograph of incoherent twin interfaces after loading dynamically 90° to the growth direction to 15.4 % strain. (b) Selected area diffraction pattern of (a). (c) Bright-field TEM micrograph after loading dynamically along the $[101]_{Ag,Cu}$ growth direction to 9.3 % strain. (d) Selected area diffraction pattern of (c).

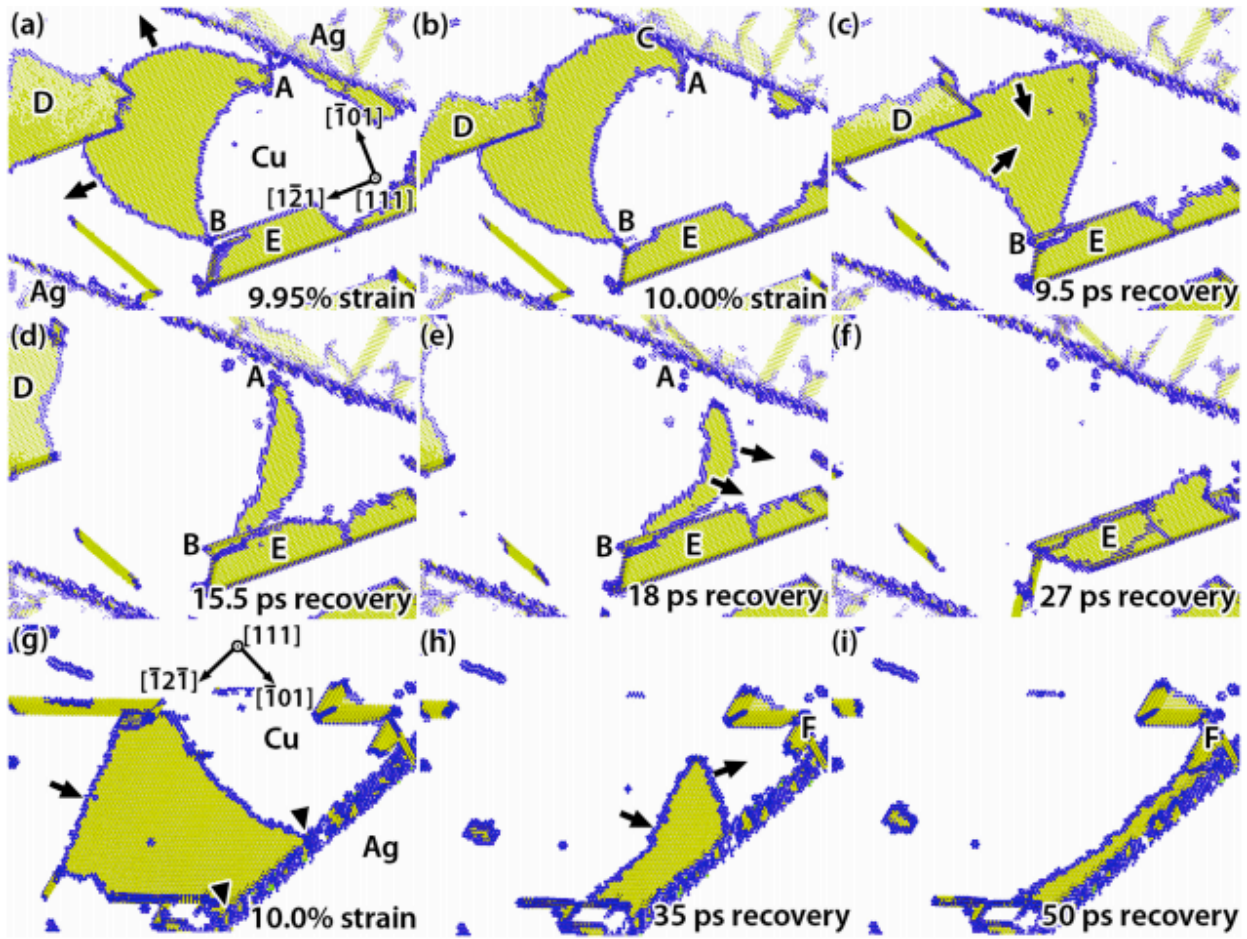

Figure 9: Dislocation mechanisms underlying plastic recovery in samples with incoherent twin interfaces. (a)-(f) Dynamics of partial dislocation mobility and pinning at the interface and dislocation tangles during loading (a)-(b) and unloading (c)-(f). A and B are two stable defect sites. D and E are complex dislocation structures. (g)-(i) Annihilation of a stacking fault and dislocation line length on retraction into the interface nucleation site. Arrows indicate the direction of dislocation motion. Arrowheads in (g) mark dislocation pinning points at the interface. In (a)-(i), blue atoms do not have local fcc, bcc, or hcp order and represent grain boundary atoms, dislocation cores, and other defect structures inside the layers. Yellow atoms have hcp structure and represent stacking faults; fcc atoms are not visualized. Recovery time

mentioned in (c)-(f) and (h)-(i) was measured from the time when the applied compressive stress was removed.

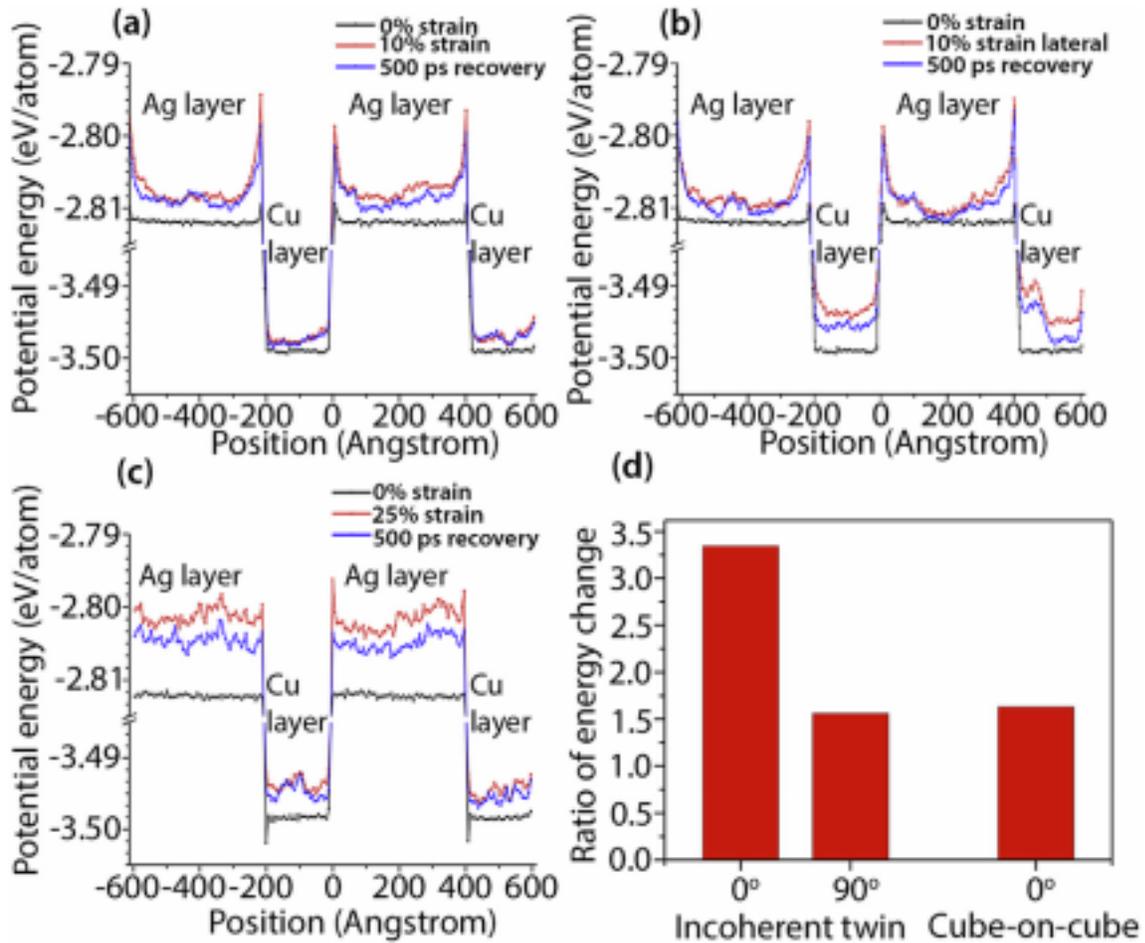

Figure 10: (a) Potential energy calculated along the direction normal to the incoherent twin interfaces in the case of loading along $[101]_{Ag}$ and $[110]_{Cu}$ for 0 % strain (before loading), 10 % strain (at the peak of loading) and at 500 ps of the recovery stage. (b) Potential energy along the direction normal to the interface for the system with incoherent twin interfaces loaded at 90° to $[101]_{Ag}$ and $[110]_{Cu}$. (c) Potential energy along the direction normal to the interface in the system with cube-on-cube interfaces loaded along $[101]_{Ag,Cu}$. (d) Ratio of energy change between the interface and the Ag layer interior.

Table 1: Compressive strain levels and strain recovered. Samples for tests 1, 2 and 6 were taken from the same directionally solidified rod and those for the other tests from a second rod. [12].

| Sample | Angle between load and $[101]_{Ag}||[110]_{Cu}$ growth direction | Strain rate $(s^{-1})$ | Strain (%) | Strain recovered (%) Elastic | Strain recovered (%) Plastic |
|---|---|---|---|---|---|
| 1 | 0° | $10^3$ | 9.3 | 0.4 | 3.2 |
| 2 (repeated loading) | 0° | $10^3$ | 10.6 | 0.4 | 2.7 |
| | 0° | $10^3$ | 15.1 | 0.4 | 2.1 |
| | 0° | $10^3$ | 17.0 | 0.4 | 1.5 |
| | 0° | $10^3$ | 21.9 | 0.4 | 0 |
| 3 | 0° | $10^3$ | 9.9 | 0.4 | 1.5 |
| 4 | 0° | $10^3$ | 6.8 | 0.4 | 1.4 |
| 5 (multiple samples) | 0° | $10^{-3}$ | 2.5-10.5 | 0.5 | 0.0 |
| 6 | 90° | $10^3$ | 15.4 | 0.5 | 0.6 |

Table 2. Compressive strain levels and strain recovered for the simulations with incoherent twin interfaces.

| Sample | Angle between load and $[101]_{Ag}\|\|[110]_{Cu}$ growth direction | Strain rate $(s^{-1})$ | Strain (%) | Strain recovered (%) Elastic | Strain recovered (%) Plastic |
|---|---|---|---|---|---|
| MD | 0° | $5 \times 10^8$ | 10.0 | 0.83 | 0.92 |
| MD | 90° | $5 \times 10^8$ | 10.0 | 1.27 | 0.77 |

Table 3: Loading conditions and change in orientation relationship. SHPB and MD stand for split-Hopkinson pressure bar and molecular dynamics, respectively.

| Test method, strain (%), and load orientation with respect to the growth direction [13] | Range of rotation away from the twin orientation relationship (average value) | Test method, strain (%), and load orientation with respect to the growth direction | Range of rotation away from the twin orientation relationship (average value) |
|---|---|---|---|
| Undeformed material | 0.7° to 2.4° (1.7°) | MD Undeformed | 2.7° to 2.9° (2.8°) |
| SHPB, 9.3 %, 0° | 5.9° and 9.8° (7.6°) | MD, 10 %, 0° | 5.5° to 7.2° (6.2°) |
| | | MD, 10 % and 500 ns recovery, 0° | 4.2° to 6.0° (5.1°) |
| SHPB, 21.9 %, 0° | 5.3° and 14.0° (10.3°) | MD, 25 %, 0° | 10.1° to 10.7° (10.4°) |
| | | MD, 25 % and 500 ns recovery, 0° | 10.2° to 10.8° (10.4°) |
| SHPB, 15.4 %, 90° | 0.0° to 5.2° (1.8°) | MD, 10 %, 90° | 2.2° to 2.9° (2.4°) |
| | | MD, 10 % and 500 ns recovery, 90° | 2.0° to 2.7° (2.3°) |
| Quasi-static, 4 %, 0° | 6.8° to 7.4° (7.0°) | | |